# Guiding Ebola Patients to Suitable Health Facilities: An SMS-based Approach

Mohamad Trad[1], Raja Jurdak[2], Rajib Rana[3]

[1] Doctors without Borders, Paris France

[2] Commonwealth Scientific Industrial and Research Organisation, Brisbane Australia

[3] University of Southern Queensland, Brisbane Australia

## Background

The current Ebola virus outbreak in West Africa that originated in Guinea in December 2013, and now has spread to two neighboring countries is not expected to be contained anytime soon. It is estimated that up to 1.4 million people may be infected by January 2015 [4]. Some of the challenges identified include the geographical spread of cases, people's movement between three countries, the transport and burial of bodies by relatives, the weak healthcare systems, severe human resource constraints, lack of proven therapeutics, and lack of community engagement and cooperation ([5][6]).

There have been numerous reports where probable Ebola-infected patients had to be driven away from health care facilities due to lack of bed availability. Furthermore, mobility of a positive case imposes further public health risks ([5][7]). Current CDC (Centre for Disease Control and Prevention) recommendations to stop the spread of Ebola include: case finding and diagnosis, case isolation and contact tracing, and safe burial practices [4].

Difficult access due to remoteness and lack of transportation [1] further widen the gap between the patient and the healthcare provider. Access to information that guides patients to the nearest facility with appropriate resources is another challenge. There is currently no known system that can improve this access.

We propose building a recommendation system based on simple SMS text messaging to help Ebola patients readily find the closest health service with available and appropriate resources. The system will map people's reported symptoms to likely Ebola case definitions and suitable health service locations. In addition to providing a valuable individual service to people with curable diseases, the proposed system will also predict population-level disease spread risk for infectious diseases using the crowdsourced symptoms from the population, as proposed by Cedric Moro (See Fig. 1)

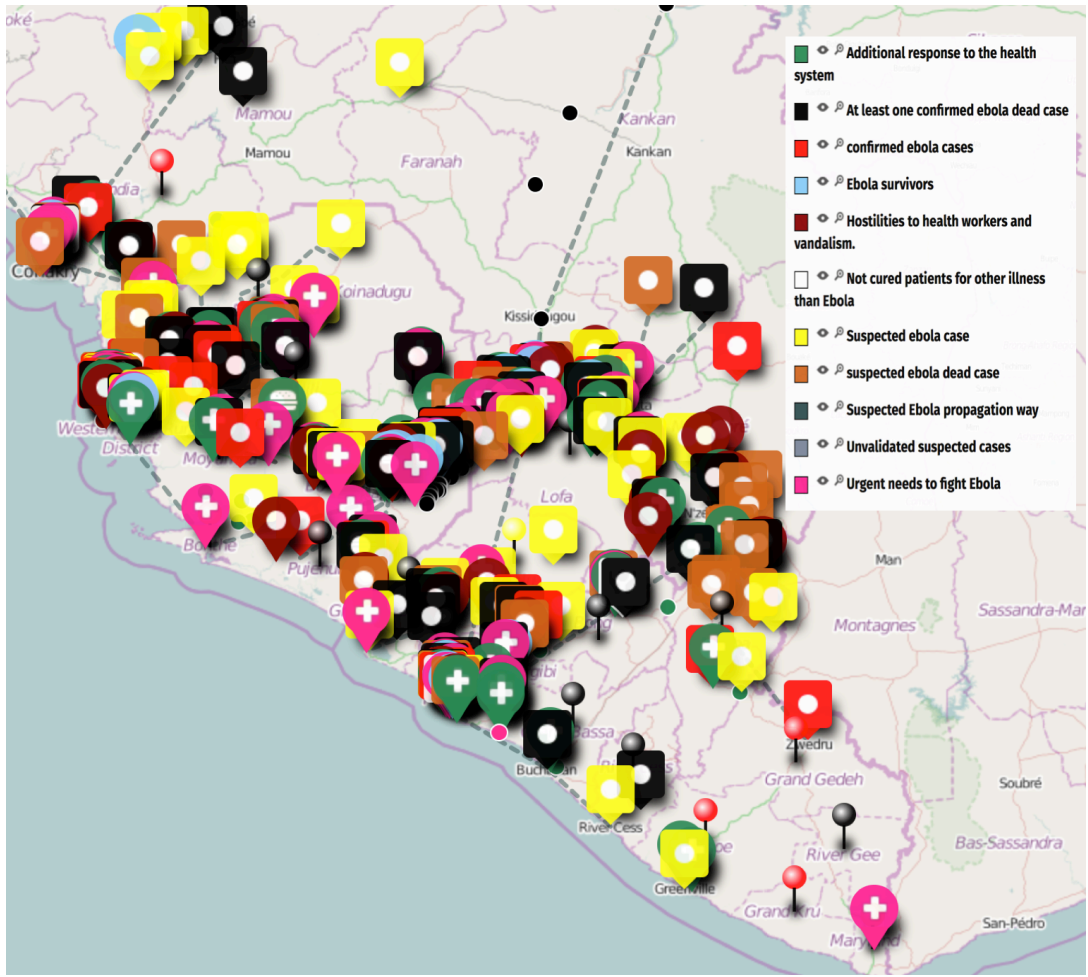

Figure 1. Ebola tracking in Sierra Leone, Liberia and Guinea by Cédric Moro - cedric.moro@i-resilience.fr

This will be extremely valuable for heath workers to better plan and anticipate responses to the current Ebola Outbreak in West Africa and hopefully prevent many new cases. The key limitation of using Cedric Moro's approach is that it only captures people using Whatsapp that requires an advanced Smartphone, so it can capture a biased sample of the population only.

In contrast, simpler and more affordable mobile phones, are highly utilized in the affected West African countries. For example, in Guinea mobile phone penetration is as high as 71 percent [2]. Therefore, mobile phones can potentially be an inexpensive yet effective way of communication in overcrowded urban or remote settings. The Somali experience with this technology has been encouraging in controlling outbreaks [3].

We propose to utilize mobile phone technology as a vehicle for people to report their symptoms and to receive immediate feedback about the health services readily available, and for predicting spatial disease outbreak risk. Once symptoms are extracted from the patient's text message, they undergo complex classification, pattern matching and prediction to recommend the nearest suitable health service. The added benefit of this approach is that it enables health care facilities to anticipate arrival of new potential Ebola cases.

**General Approach**

The workflow of the proposed end-to-end system is shown in Fig. 2. The use case for the patients will be straightforward. Using any basic mobile phone they can send a text message about their symptoms based on CDC case definition [4] to a designated toll- free number. The message once received will be analyzed with Natural Language Processing Algorithms to extract the symptoms. Patient location will be approximated by the available cell-tower information or provided post-code or village name. Symptoms will then be fed to the classification module to determine whether it is a likely Ebola case and to find the nearest health service with sufficient resources to treat the disease. The classification module will use CDC provided clinical mapping of symptoms to Ebola as the foundation for classification. A limitation would be to correctly classify an Ebola case with high accuracy and precision with minimal false classifications, as many differentials share the symptoms of Ebola virus infection. Our classification module will also take into account available meta information, such as the spatial extent of the current outbreak to accurately classify the disease. In addition, a user feedback mechanism will be in place for fine-tuning the classification module.

To obtain information about currently available facilities in a health service, we will adopt a phased approach. The knowledgebase will be initially constructed using the location and contact number of the health services extracted from web portals using an automated web crawler; later on, we will use both positive and negative patient feedback to infer the current facilities at the service. For example, if patient feedback shows the recommendation was useful, we will assume that the health service has capacity to admit more patients. Otherwise, we will assume that the health service is out of capacity and keep the status for some time (need to determine). However, there will be a provision for the health services to update their information anytime regarding resource, location (e.g. NGO field clinics) in the knowledgebase using text messages. At the server end the text messages indicating information update will undergo Natural Language Processing algorithms to extract the key words (e.g., how many beds available) and the knowledge base will be updated accordingly. Once the information is updated, the health services/patients will be notified using a reply text message.

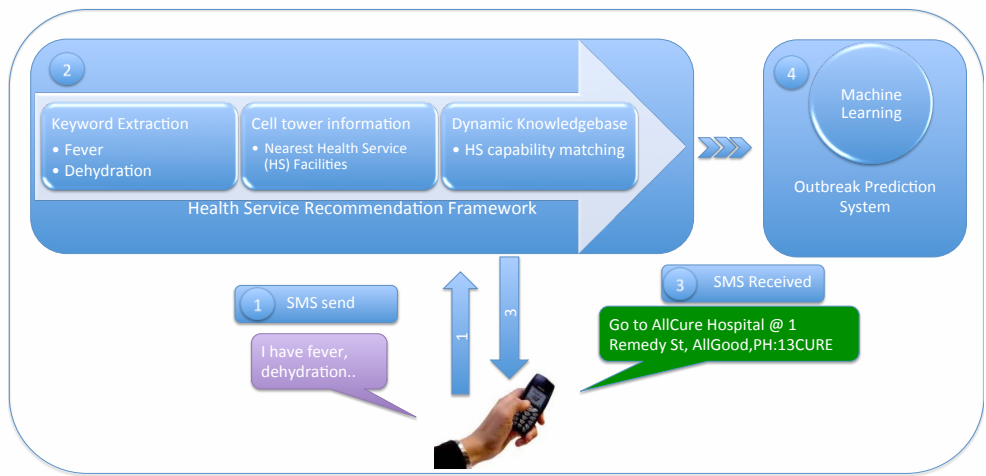

Figure 2. Recommendation Framework.

The text messages received from the population of end-users will be continuously analyzed at the back-end to predict the outbreak risk into the future and its spread pathways. The locations of users extracted from the text messages will be used to estimate their orbits of movement. These locations will be coupled with the inferred patient conditions to predict the geographic extent of disease-spread risk.

The current Ebola outbreak is shaping up to be a large scale problem, not only in the 3 affected West African countries, but also with cases starting to appear in Western countries such as Spain and the USA. The CDC recommendations for controlling the spread further as at the source, i.e. in West Africa. Given the large spatial scale of the disease, any assistive technology should be equally prevalent and widely used in the target countries. It is particularly for this reason that we believe and advocate for the use of simple SMS messaging as a mechanism to guide Ebola patients to suitable health facilities.

## 2. Investment Outcomes

We request a total of 190,000 AUD to execute this project (1 FTEs per year @ 250,000 to cover software engineering and research hire). The intended investment outcomes and budget justification is the following table.

| Primary Outcome/Results | Results Measurement | Resource/Time |
|---|---|---|
| 1. Developing a health service recommendation software framework | A comprehensive simulation system will be developed to validate the recommendation framework. Simulated patients based on empirical statistics on reported symptoms in the population will send text messages and recommendation will be made based on the symptoms. The performance of the Natural Language Processing algorithms will be measured as the first step to accurately extract the keywords from the text messages. As the second step, the simulation system will facilitate rigorous manifold cross validation of the classification module and will quantify its accuracy. Finally, the quality of advice (i.e. accuracy) will be measured quantitatively based on the satisfaction feedback of simulated patients. | 1 FTE of research labor for one year will be used to deliver this outcome. |
| 2. Developing Outbreak prediction model | Our system will simulate regions with varying demographic profiles and their interconnectivity will be modeled using graph theory and statistical physics. We will investigate the feasibility of various diffusion models, for instance | 1 FTE of a research labor time will be used |

## References

[1] Sheik - Mohamed, Abdikarim, and Johan P. Velema. "Where health care has no access: the nomadic populations of sub‐Saharan Africa." Tropical medicine & international health 4.10 (1999): 695-707.

[2] http://www.budde.com.au/Research/Guinea-Telecoms-Mobile-and-Broadband-Market-Insights-Statistics.html

[3] http://blogs.oxfam.org/en/blogs/12-11-13-mobile-phones-help-tackle-cholera-somalia

[4] (http://www.cdc.gov/vhf/ebola/resources/index.html)

[5] Lancet, The. "Ebola in west Africa: gaining community trust and confidence." The Lancet 383, no. 9933 (2014): 1946.

[6] Trad, Mohamad-Ali, Dale Andrew Fisher, and Paul Anantharajah Tambyah. "Ebola in west Africa." The Lancet Infectious Diseases (2014).

[7] Médecins Sans Frontières (MSF), http://www.msf.org